# Red-edge position of habitable exoplanets around M-dwarfs

Kenji Takizawa[1,2], Jun Minagawa[2], Motohide Tamura[1,3,4], Nobuhiko Kusakabe[1,4], *Norio Narita[1,3,4]

[1] Astrobiology Center, National Institutes of Natural Sciences, 2-21-1 Osawa, Mitaka, Tokyo 181-8588, Japan

[2] National Institute for Basic Biology, National Institutes of Natural Sciences, 38 Nishigonaka, Myodaiji, Okazaki, Aichi 444-8585, Japan

[3] Department of Astronomy, The University of Tokyo, 7-3-1 Hongo, Bunkyo-ku, Tokyo 113-0033, Japan

[4] National Astronomical Observatory of Japan, National Institutes of Natural Sciences, 2-21-1 Osawa, Mitaka, Tokyo 181-8588, Japan

*Corresponding author: Norio Narita (norio.narita@nao.ac.jp)

**Abstract**

One of the possible signs of life on distant habitable exoplanets is the red-edge, which is a rise in the reflectivity of planets between visible and near-infrared (NIR) wavelengths. Previous studies suggested the possibility that the red-edge position for habitable exoplanets around M dwarfs may be shifted to a longer wavelength than that for Earth. We investigated plausible red-edge position in terms of the light environment during the course of the evolution of phototrophs. We show that phototrophs on M-dwarf habitable exoplanets may use visible light when they first evolve in the ocean and when they first colonize the land. The adaptive evolution of oxygenic photosynthesis may eventually also use NIR radiation, by one of two photochemical reaction centers, with the other center continuing to use visible light. These "two-color" reaction centers can absorb more photons, but they will encounter difficulty in adapting to drastically changing light conditions at the boundary between land and water. NIR photosynthesis can be more productive on land, though its evolution would be preceded by the Earth-type vegetation. Thus, the red-edge position caused by photosynthetic organisms on habitable M-dwarf exoplanets could initially be similar to that on Earth and later move to a longer wavelength.

**Introduction**

According to recent observations by NASA's Kepler mission, it has become evident that habitable Earth-like exoplanets are widely distributed throughout the universe.[1,2] If a number of such exoplanets are discovered around solar neighborhood stars by near-term space missions, such as the Transiting Exoplanet Survey Satellite (TESS),[3] and ground-based precision radial velocity surveys, searches for signs of life (i.e., biosignatures) would become a reality for big scientific projects. Encouragingly, it has been suggested that the nearest star Proxima Centauri has a habitable Earth-mass planet.[4] In fact, searching for biosignatures on habitable exoplanets is an explicit goal of many near-term space survey missions, including next-generation 30-meter class ground-based telescopes such as the European Extremely Large Telescope, the Thirty Meter Telescope, and the Giant Magellan Telescope, and NASA's 2020 decadal survey space missions, such as the Habitable-Exoplanet Imaging Mission and Large UV/Optical/Infrared Surveyor. Thus, it is important to study the validity of possible biosignatures to appropriately design future missions.

Oxygen is one of the most discussed biosignature candidates.[5,6] This is reasonable because two essential sources for oxygenic photosynthesis—light and liquid water—should by definition be available on habitable exoplanets, and oxygenic phototrophs can significantly change atmospheric compositions, as proven on Earth.[7] However, recent studies have shown that oxygen can also be a false positive sign of life on habitable exoplanets.[8-11] Thus, additional biosignatures will be necessary to determine whether oxygen is produced by phototrophs or abiotic processes. One possible additional biosignature for oxygenic phototrophs is the red-edge, which is a rise in a planet's surface reflectivity between red absorbance and near-infrared (NIR) reflection due to vegetation.[12,13] Although the red-edge position for Earth's vegetation is fixed at around 700–760 nm, that for exoplanets may not necessarily be the same.[14-16] Thus, it is necessary to predict the red-edge wavelength position in advance in order to use it as an additional biosignature for distant habitable exoplanets.

Previous studies proposed that the red-edge position may be shifted toward longer wavelength if three- or four-photon photosynthesis is expressed on habitable exoplanets, especially around M dwarfs.[14,15,17] M dwarfs have lower surface temperatures (2,500–3,800 K) than the Sun (5,777 K) and emit much more strongly at NIR wavelengths than at visible wavelengths. Hence, the possibility of a shifted red-edge position may be reasonable if oxygenic phototrophs evolve to use the most abundant photons for photosynthesis. While NIR light is abundant on land, only visible light penetrates more than 1 m into the water column[14,17] (see the area graph in Figure 1). It is predicted that Earth-type marine phototrophs can grow using visible light in an M-dwarf planet's ocean, where phototrophs can be protected from ultraviolet (UV) flares from young M dwarfs.[14] Adaptations for land surface and underwater were described independently in previous studies[14,17], but evolution from the water to the land (i.e., adaptation for a transient environment) has remained unexplored. Without solving this missing link, NIR-light photosynthesis and a shifted red-edge position cannot be an effective prediction.

Recently, the number of possible habitable exoplanets discovered around nearby M dwarfs has been increasing.[4,18] Indeed, 4 Earth-like exoplanets

TRAPPIST-1 e,f,g [19] and LHS 1140 b [20] are recently discovered to be orbiting around nearby M dwarfs and within the conservative habitable zone [21]. Thus, now is the time to re-examine the red-edge position of habitable M-dwarf exoplanets. In this paper, as a first step, we predict photosynthetic machineries that could emerge on land and underwater as a result of adaptive evolution as in previous works [14,17]. Then, for the first time, we examine whether the hypothetical land phototrophs can evolve from marine phototrophs (as conceptualized in the right-hand panel in Figure 1).

**Results**

- Light environment on habitable planets around M-dwarf stars

Light conditions on habitable planets orbiting different types of stars were studied previously by Walftencroft and Raven[17] and more extensively by Kiang et al.[14] Here, we recalculated M-dwarf radiation on habitable exoplanets and obtained results almost identical to those of the previous studies. AD Leonis (AD Leo), a dM3e star located just 16 light years away from our solar system[5] was chosen as a model case. We assumed that an Earth-like planet is located in the habitable zone around AD Leo and that it receives the same amount of radiation energy as present-day Earth, so that light conditions on the planet can be estimated and compared with solar irradiation on the Earth.

Table 1 summarizes the parameters representing the light conditions on habitable planets around the Sun and AD Leo. The energy flux density (EFD) spectra and photon flux density (PFD) spectra (Supplemental Figure 1) on the Earth's surface peak at visible wavelengths, while these peaks are red-shifted to around 1,000 nm under AD Leo's radiation. The intensity of photosynthetically active radiation can be estimated by integrating the PFD within a given wavelength range. When the absorption bandwidth is fixed at 300 nm, the integrated PFD is at a maximum at 578−878 nm on Earth and 985−1,285 nm under AD Leo's radiation (Supplementary Figure 2a,b). The absorption wavelength range for terrestrial green plants (400−700 nm) is close to the optimal range for solar radiation (integrated PFD is 88% of the maximum), but shifted from the optimal position under AD Leo (12% of the maximum). Assuming that all photons in the absorption range are funneled into the reaction centers and converted into excitation energy equivalent to the longest-wavelength photons, the total excitation energy obtained on the Earth is at a maximum when the absorption range is 447−747 nm (Supplementary Figure 3a). The total excitation energy obtained by green plants (absorbing at 400−700 nm) is 98% of the maximum value, which is optimal for solar radiation. If the same calculation is applied to AD Leo's radiation, the optimal absorption range is around 900−1,100 nm and the suboptimal range (70% of the maximum energy gain) is around 700−900 nm (Supplemental Figure 3b). It is the simplest estimation of photosynthetically active radiation from an M-dwarf star, though there is no reason to fix the absorption bandwidth at 300 nm except that this is the case for plants on Earth. If there is no limitation, the integrated PFD increases as the absorption bandwidth expands (Supplemental Figure 2c,d). To maximize the energy gain at reaction centers, the optimal absorption range is around 400−1,100 nm and 400−1,300 nm for solar and AD Leo radiation, respectively. Although there should be a limit to the use of longer-wavelength photons, phototrophs could evolve to use NIR radiation on the land surface of an M-dwarf planet and even on the Earth.

The underwater lighting conditions are very different from that on land because NIR radiation is strongly attenuated by water molecules.[22] All optical parameters (peak wavelength in EFD and PFD spectra, integrated PFD with fixed and open bandwidth, and total energy gain at reaction centers) suggest that visible light is optimal for marine phototrophs on both planets. The visible light intensity in the ocean of AD Leo planet is about 10% of that for the present Earth. Assuming that oxygenic photosynthesis evolved about 2.7 billion years ago [23,24] when the sun was fainter than today, the light intensity in the ocean of AD Leo planet is about 15% of that for the Earth (Supplemental Table 1). In general, Earth-type phototrophs would be less active under M-dwarf radiation, though some Earth marine phototrophs, which can grow under low light conditions, may find an appropriate habitat on M-dwarf planets, as predicted in previous studies.[14,17]

- Utilization of NIR radiation for photosynthesis

There is no doubt that the use of NIR radiation for photosynthesis would be a big advantage for phototrophs on the land surface of habitable planets around M dwarfs. On the other hand, Earth-type oxygenic photosynthesis may evolve in water at depths below 1 m before ozone layer formation. In this section, we examine the hypothetical NIR-using phototrophs as descendants of Earth-type phototrophs.

Before discussing possible photosynthetic mechanisms expected around M dwarfs, we will briefly review photosynthesis on Earth. Oxygenic photosynthesis is a combined reaction of water oxidation and $CO_2$ reduction to form a biologically useful high-energy compound, carbohydrate. The individual half reactions and their standard redox potential at pH 7.0 ($E^{0'}$) are as follows:

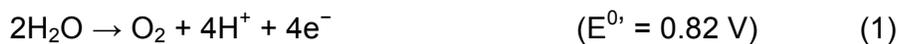
$$2H_2O \rightarrow O_2 + 4H^+ + 4e^- \qquad (E^{0'} = 0.82 \text{ V}) \qquad (1)$$

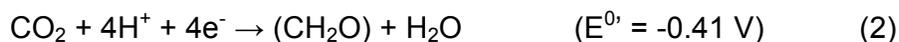
$$CO_2 + 4H^+ + 4e^- \rightarrow (CH_2O) + H_2O \qquad (E^{0'} = -0.41 \text{ V}) \qquad (2)$$

Since the electron transfer reaction is energetically uphill ($\Delta E^{0'} = -1.23$ V), light energy input is required for the reaction to occur. Earth-type photosynthesis consists of two photochemical reaction centers driving the liner electron flow (LEF) from $H_2O$ to $CO_2$ through two-step sequential light excitation (two-photon reaction). The excitation wavelengths for the reaction center pigments are 680 and 700 nm, and these pigments are therefore named P680 and P700, respectively. The reaction centers are surrounded by a variety of light-harvesting pigments that have the same or shorter excitation wavelengths.

Photosynthetically useful radiation at wavelengths from 400 to 700 nm is captured by light-harvesting pigments, and excitation energy is funneled into the reaction centers. For simplicity, in this section, except when it is specifically mentioned, we considered that the two Earth-type reaction centers have the same excitation wavelength of 700 nm and that four times of two-photon reactions occur (eight photochemical reactions in total), producing one $O_2$ molecule. Two photons (or excitation energy) at 700 nm correspond to 3.54 V, which is significantly larger than the minimum energy requirement of 1.23 V. The simplified model predicts an energy conversion efficiency of 35%, as reported by Hill and Rich.[25] The large amount of redox energy (65%) is used as a driving force of electron transfer reactions and thus enables nearly 100% quantum efficiency under optimal conditions.

Oxygenic photosynthesis could be driven by low-energy NIR radiation by improving the redox energy conversion efficiency and/or employing multi-photon reactions. The longest known wavelength of photochemical excitation that can be used for oxygenic photosynthesis on the Earth is found in chlorophyll d-containing cyanobacteria, where two reaction centers absorb at around 715 and 740 nm.[26] It was revealed that the energy conversion efficiency of chlorophyll d-using light reactions is improved by about 5% without decreasing quantum efficiency (i.e. without decreasing oxygen/photon ratio) compared to chlorophyll a-using reactions.[27] We assume that the upper limit for energy conversion efficiency is about 40% in Earth-type photosynthesis (it could be exceeded in future and/or on other planets). Further improvement in energy conversion efficiency may decrease thermodynamic driving force for the appropriate electron transfer reactions and thus increase the risk of harmful side reactions such as production of reactive oxygen species.[28,29] Such a small red shift in the reaction center wavelength does not have a significant effect on estimating the red-edge position for exoplanets, though it is important for the organism to adapt to the particular environment.

Another popular hypothesis involves a three- or four-photon reaction, instead of a two-photon reaction. If an efficiency of 35% is applicable to the three- and four-photon models, the excitation wavelengths are then 1,050 and 1,400 nm, respectively. Considering the reaction scheme, however, the energy conversion efficiency for a multi-photon reaction can be lower than 35%. In the two-photon model, 65% of the excitation energy (2.31 V) is lost as heat, mainly within ~20 sequential electron transfer reaction steps. Assuming that reaction steps in three- and four-photon models increase 1.5 and 2 times when compared to the two-photon model, the energy losses (i.e., the driving force for electron transfer reactions) during the reaction steps will increase to 3.47 and 4.62 V, respectively. In these cases, the corresponding wavelengths are 791 nm for the three-photon model and 847 nm for the four-photon model (see Supplemental Figure 4). Considering that several reaction steps (such as water oxidation) are common to all the models, this assumption may overestimate the energy loss. The appropriate wavelength for multiple-photon models may be somewhere between the two estimates: 900 nm for the three-photon model and 1,100 nm for the four-photon model. If we calculate the $O_2$ production rate under AD Leo radiation, the two-photon reaction using 400−700-nm radiation, the three-photon reaction using 400−900 nm radiation, and the four-photon reaction using 400−1,100 nm radiation can produce $O_2$ at 15%, 51% and 80%, respectively, of the land plant $O_2$ production rate on Earth (Supplemental Figure 5). The productivity of the multi-photon reactions is slightly less than that in previous estimates[14,17] but is still significantly higher than that for a two-photon reaction using visible light.

If two (or multiple) reaction centers are replaced one by one during adaptive evolution, transient phototrophs may have "two-color" reaction centers: one reaction center using visible radiation and another using NIR radiation. In addition to being a transient model, the two-color reaction center is an attractive adaptation hypothesis, since it can use both high-energy visible radiation and abundant NIR radiation, as proposed by Blankenship et al.[30] Assuming that one reaction center is excited by visible radiation at 700 nm and the other excited by NIR at 900 nm, a two-photon reaction can produce enough energy for oxygenic photosynthesis at an overall energy conversion efficiency of 40%.

- Transition from water to land

In the previous sections, we studied the lighting conditions and photosynthetic apparatus under two different environments, on land and underwater. Here, we will examine intermediate lighting conditions in shallow water and how phototrophs might adapt to it. Figure 2 shows the photosynthetically active radiation intensities received by phototrophs when they move from the water surface to 100 m below the surface (active movement by phototaxis or passive movement by vertical mixing). The PFDs for the sun and AD Leo (Supplementary Figure 1) were integrated in the ranges of 400–700, 400–900, and 400–1,100 nm and are plotted as a function of water depth. Under solar radiation (Figure 2a), the integrated PFD for 400–700 nm increases 1.4 times when the water depth decreases from 1 to 0 m. A relatively small change in the effective light intensity allows for a prompt transition from an aquatic phototroph to a land phototroph without the need to change its light-harvesting apparatus.

Under AD Leo's radiation (Figure 2b), the change in the integrated PFD for 400–700 nm is 1.7 times, suggesting a relatively larger light stress for Earth-type phototrophs. Under the same conditions, the integrated PFDs for 400–900 and 400–1,100 nm increase 8 and 16 times, respectively. In addition to the change in light intensity, the spectral change is large in shallow water (Figure 1). Therefore, if phototrophs using NIR radiation evolve underwater, they will be exposed to strong light intensity and quality changes when approaching the water surface. This may be a particularly big hindrance for the two-color reaction, which has to balance light excitation between visible and NIR radiation using reaction centers under a widely changing light spectrum.

In order to maintain excitation balances, land plants and aquatic algae on Earth have a dynamic antenna regulation mechanism called state transitions.[31-34] State transitions were discovered as a phenomenon to rebalance light distribution to P680 and P700 when either one is preferentially excited.[35,36] The molecular mechanism involves a phosphorylation of light-harvesting protein depending on the redox state of a mobile electron carrier between P680 and P700 and reversible movement of the phosphorylated antenna between two reaction centers.[37] While the fraction of mobile antennae varies depending with species, green algae live in shallow water and possess the largest capacity (about 40% of P680 antenna for the highest estimation[32,33]). This may be one of the reasons why green algae are well adapted to shallow water and land plants evolved from green algae.[38,39] Such a state transition mechanism would be necessary on habitable exoplanets.

Hereafter, we examine whether two-color photosynthesis can adapt to the changing light conditions via state transitions. As a working model, we predicted the biphasic state transitions as shown by Figure 3. The visible light-using reaction center ($P_{vis}$) and NIR-using reaction center ($P_{NIR}$) are associated with specific light-harvesting antennae. The $P_{vis}$ antenna absorbs visible radiation, while the $P_{NIR}$ antenna absorbs visible and NIR radiation. In the initial state, a photosynthetic cell is placed underwater at a certain depth, and the antenna size for the two reaction centers is optimized for the given light conditions. When the cell descends, and NIR radiation decreases more than visible radiation (see Figure 1), a part of the $P_{vis}$ antenna moves from $P_{vis}$ to $P_{NIR}$ to redistribute the excitation energy[33] (referred to as the shuttling model). When the cell rises, and both visible and NIR light exceed the optimal intensity (i.e., energy input exceeds energy demand), a part of the $P_{vis}$ antenna detaches from $P_{vis}$ and works as a quencher for dissipating excess energy safely[34] (the quenching model). Excess energy at $P_{NIR}$ is considered to be used for cyclic electron flow[40,41] (CEF) around $P_{NIR}$, which dissipates excitation energy without net production of reductants (we simply estimated required activity of CEF for complimenting the regulation by state transitions without considering its own regulatory roles). Using this model, we calculated the size of the moving antenna and resulting electron flow under changing light conditions.

To simplify the calculations, the reaction centers were assumed to be homogeneously distributed and the molecular density of the reaction center ($n$) and light path length ($l$) were held constant. Under these conditions, the relative change in absorption cross section ($\sigma$) is equal to the relative change in absorbance (Abs) from the Beer-Lambert law: Abs=$\log_{10}(e)*\sigma*n*l$. The light excitation frequency for $P_{vis}$ (ExP$_{vis}$), and $P_{NIR}$ (ExP$_{NIR}$) were determined by the given light intensity at visible ($I_{vis}$) and NIR ($I_{NIR}$) wavelengths, the ratio of the light intensity captured by each antenna type, and the fraction of the antennae associated with the reaction centers as

$$\text{ExP}_{vis} = I_{vis} * (1 - 10^{-(\alpha+\beta)}) * \alpha/(\alpha+\beta) \tag{3}$$

$$\text{ExP}_{NIR} = I_{vis} * (1 - 10^{-(\alpha+\beta)}) * \beta/(\alpha+\beta) + I_{NIR} * (1 - 10^{-\beta}), \tag{4}$$

where α and β are the antenna size (Abs) for $P_{vis}$ and $P_{NIR}$, respectively. In the initial state, the antenna sizes are optimized to balance the excitation frequencies. If the light intensities ($I_{vis}$ and $I_{NIR}$) and the total antenna size ($α_i+β_i$) are given, the initial antenna size ($α_i$ and $β_i$) can be calculated by solving the equation ExP$_{vis}$=ExP$_{NIR}$.

When the cell descends, a part of the $P_{vis}$ antenna (size at α') moves to decrease the antenna size associated with $P_{vis}$ ($α_i$-α') and increase the antenna size associated with $P_{NIR}$ ($β_i$+α'). The total antenna size remains constant ($α_i+β_i$). In this case, the excitation frequencies are calculated as

$$\text{ExP}_{vis} = I_{vis} * (1 - 10^{-(\alpha i+\beta i)}) * (\alpha_i - \alpha')/(\alpha_i + \beta_i) \tag{5}$$

$$\text{ExP}_{NIR} = I_{vis} * (1 - 10^{-(\alpha i+\beta i)}) * (\beta_i + \alpha')/(\alpha_i + \beta_i) + I_{NIR} * (1 - 10^{-\beta i}). \tag{6}$$

Since the shuttling model keeps the excitation in balance, α' is calculated by solving the equation ExP$_{vis}$=ExP$_{NIR}$. Under equal excitation of reaction centers, all excitation energy is used for LEF (LEF=ExP$_{vis}$=ExP$_{NIR}$).

When the cell rises and visible radiation increases, the mobile fraction of $P_{vis}$ antenna (α') is in a quenching state. It decreases the antenna size for $P_{vis}$ ($α_i$-α') without changing the antenna size for $P_{NIR}$ ($β_i$) and the total antenna size ($α_i+β_i$). In this case, the excitation frequencies are calculated as

$$\mathrm{ExP_{vis}} = I_{vis} * \left(1 - 10^{-(\alpha i + \beta i)}\right) * (\alpha_i - \alpha')/(\alpha_i + \beta_i) \qquad (7)$$

$$\mathrm{ExP_{NIR}} = I_{vis} * \left(1 - 10^{-(\alpha i + \beta i)}\right) * \beta_i/(\alpha_i + \beta_i) + I_{NIR} * \left(1 - 10^{-\beta i}\right). \qquad (8)$$

The quenching model keeps ExP$_{vis}$ and LEF in the initial state regardless of increasing light intensity; therefore, α' is calculated by substituting the initial value of ExP$_{vis}$ into equation 7. ExP$_{NIR}$ is calculated as a linear function of the light intensity by substituting the initial antenna sizes in equation 8. Excessive excitation energy at P$_{NIR}$ is used for CEF (CEF=ExP$_{NIR}$-ExP$_{vis}$).

Before using it for exoplanet photosynthesis, we confirmed the state transition model by applying it to Earth-type photosynthesis under solar radiation. The light excitation frequencies for P680 and P700 were calculated instead of P$_{vis}$ and P$_{NIR}$, assuming that the P680 antenna and P700 antenna absorb at 400−680 nm and 400−700 nm, respectively, and the total antenna size is 1.0 Abs (absorbing 90% of radiation). The light intensities at wavelength of 400−680 nm and 680−700 nm (corresponding to $I_{vis}$ and $I_{NIR}$) were calculated by integrating the PFD spectra (Supplementary Figure 1a). When the cell was placed at the water surface, the excitation frequency for P680 and P700 (calculated by equations 3 and 4) were balanced when the P680 and P700 antenna sizes (α$_i$ and β$_i$) were 0.53 and 0.47, respectively. When the surface-acclimated cell descends, antenna movement from P680 to P700 (α' in equations 5 and 6) changes the antenna sizes for P680 and P700 to 0.50 and 0.50 (6% of the P680 antenna is shuttling) for keeping the light excitation in balance (Figure 4a). When the cell is acclimated at a depth of 1 m, the antenna sizes for P680 and P700 are 0.51 and 0.49, respectively, for balancing light excitation. When the cell rises, a part of the P680 antenna (α' in equation 7 and 8) transitions to a quenching state. The P680 antenna size is decreased to 0.38 (25% of the P680 antenna is in a quenching state) as the cell approaches the surface, while the P700 antenna size remains at 0.49. An increase in light intensity increases P700 excitation (calculated by equation 8) and activates CEF up to 44% of LEF (31% of the maximum LEF for the surface-adapted cells), as shown in Figure 5b. When the cell descends after acclimated at a depth of 1 m, P680 antenna moves from P680 to P700 as in the surface-acclimated cell but the fraction of mobile antenna are smaller (2% of the P680 antenna is shuttling). We use the surface-acclimated cell for evaluating the antenna regulation capacity by the shuttling model. Light acclimation at deeper than 1 m is not considered since NIR radiation is negligible. The flexibility of the antenna size and the CEF/LEF response demonstrated in the model calculations are within the capabilities of actual green algae and plants shown by *in vivo* experiments.[32,33,42,43]

In the next step, the state transitions for Earth-type photosynthesis were calculated under AD Leo radiation. The light intensities at wavelengths of 400−680 and 680−700 nm under AD Leo radiation were calculated from the PFD spectra (Supplementary Figure 1b). Other conditions were identical to those for the calculations under solar radiation. When the cell is acclimated at the water surface and at a depth of 1 m, P680/P700 antenna sizes are 0.56/0.44 and 0.53/0.47, respectively. When the surface-acclimated cell descends, the antenna sizes for P680 and P700 both become 0.50 via shuttling of 10% of the P680 antenna (Figure 4b). When the cell acclimated at 1 m depth rises to the surface, 35% of the P680 antenna is in the quenching state and the P680 antenna size decreases to 0.34 (Figure 4e). During upward cell movement, the P700 antenna size remains constant at 0.47 and CEF increases to 73% of LEF (Figure 5e). The results show that greater flexibility is required for antenna regulation and CEF/LEF activation under M-dwarf radiation compared to solar radiation.

Finally, we examined whether state transitions can adapt the hypothetical two-color reaction to the changing light conditions in shallow water on the M-dwarf planet. The excitation wavelengths for the reaction centers were assumed to be 700 and 900 nm (P700 and P900). The absorbing wavelengths for P700 and P900 antennae were 400−700 and 400−900 nm, respectively. The light intensities at 400−700 nm ($I_{vis}$) and 700−900 nm ($I_{NIR}$) were calculated from the PFD spectra (Supplementary Figure 1b). Other conditions were the same as for the previous model calculations. When the cell is acclimated with AD Leo radiation at the water surface, the sizes of the P700 and P900 antennae are 0.91 and 0.09, respectively, for balancing light excitation. The large deviation in the antenna size reflects the high $I_{NIR}$ at the water surface. As the cell descends in the water, $I_{NIR}$ decreases; therefore, the antenna sizes for P700 and P900 both become 0.50 via 45% of the P700 antenna (41% of the total antenna) shuttling from P700 to P900 (Figure 4c). At the water surface, the LEF for the two-color reaction is 1.9 times higher than for the Earth-type reaction (compare Figures 5b and 5c), confirming higher productivity by using NIR radiation in addition to visible radiation. When the two-color reaction cell is placed underwater at 1 m depth, the P700 and P900 antenna sizes are 0.52 and 0.48, respectively. As the cell rises in the water, the P700 antenna size decreases to 0.31 (Figure 4f) for keeping LEF at the acclimated level (Figure 5f), and 40% of the P700 antenna (21 % of the total antenna) is in the quenching state for dissipating excess $I_{vis}$. The P900 antenna size remains at 0.48. As $I_{NIR}$ increases when the cell approaches the water surface, P900 excitation increases correspondingly and activates CEF. CEF is 9.9 times higher than LEF (3.4 times higher than the maximum LEF) at the water surface (Figure 5f).

These results suggest that very great flexibility is required for antenna regulation and CEF activation to adapt two-color photosynthesis to M-dwarf radiation. Reversible regulation of 40% of the total light harvesting antenna could be exceeds the capacity of state transitions inherent in Earth-type photosynthesis.[32,33] Moreover, very large CEF activity exceeding maximum LEF has not been explained by the regulation mechanism in Earth-type photosynthesis.[42,43] Since a change in the CEF/LEF ratio changes the ATP/NADPH production ratio,[44] an additional mechanism to down-regulate P900 excitation, down-regulate ATP production, and/or up-regulate ATP consumption would be required for two-color reaction cells in order to adapt to the light conditions in shallow water on an M-dwarf planet.

**Discussion**

The results of our estimations are summarized as follows: (1) adaptive evolution of oxygenic photosynthesis may develop the ability to use NIR radiation on land and visible light in the ocean, as predicted by previous studies; (2) phototrophs using NIR radiation are exposed to a broad light spectrum and intensity changes in shallow water during water-to-land evolution; and (3) a pair of reaction centers with one utilizing NIR radiation requires extremely flexible antenna regulation mechanisms and additional regulatory mechanisms for adapting to a variable light environment.

After comparative investigations of light environments on the hypothetical habitable exoplanet around AD Leo and on the Earth, we conclude that two-photon reactions using visible radiation may first evolve in an M-dwarf planet's ocean as oxygenic phototrophs. Reactions using NIR radiation may then evolve on the land surface. The first oxygenic phototroph is most likely to be established underwater about 10 m[14] or deeper, and may expand its habitat to shallow water after the formation of an ozone layer and/or the cessation of UV emission from the active M dwarf. At a depth of 1 m or less, the abundance of NIR radiation may stimulate the evolution of phototrophs to use it. One of the plausible evolutionary paths from visible-radiation to NIR-radiation photosynthesis is a "two-color reaction," which consists of different reaction centers using visible and NIR radiation. However, phototrophs using NIR radiation are exposed to drastically changing light conditions in shallow water, which may be an obstacle to using NIR radiation before land colonization. The formation of the ozone layer on the Earth prior to the emergence of land plants enabled them to quickly colonize the land, safe from the effects of UV. A rapid transition from aquatic algae to land plants was accomplished without a change in the photosynthetic machinery.[39]

A seamless transition from water to land could be achieved if three factors were in place: a small difference in the radiation spectrum, a small difference in the light excitation spectra of the reaction centers, and a flexible antenna regulation mechanism to balance light excitation of the reaction centers. Although the difference in the radiation spectrum is larger on habitable exoplanets around M dwarfs, the same evolutionary path may also occur if the antenna regulation mechanism is effective. Our model calculations showed that the capacity of state transitions in green algae is large enough to adapt to M-dwarf radiation. Land plants using visible radiation can flourish under M-dwarf radiation, though they are less productive than phototrophs using NIR radiation. We propose that the first land plants on M-dwarf planets may use visible light and flourish for a significant period without using NIR radiation. These land plants should exhibit a red-edge position at 700–760 nm, as on the Earth. When the red-edge is used as a biosignature for extrasolar life surveys, this wavelength range should be included as one of the main targets, even for exoplanets located around M dwarfs.

Evolution to using NIR radiation could be easier after land colonization, rather than in the water. On the land surface, where the radiation spectral change is relatively small, phototrophs using NIR radiation can overcome competition with ancestral phototrophs that use visible radiation in the long run. A two-color reaction using wavelengths up to 900 nm could be the first candidate, since it can be accomplished by minor modifications of a single reaction center and electron transfer reactions around it.[30] More productive three- or four-photon reactions using wavelengths of 1,100 nm or longer are theoretically possible, but there are no known examples of these types of photosynthesis on the Earth. Grate diversity in construction of core proteins for reaction centers revealed from recent genetic analyses [45,46] is supporting the possibility that three or four different reaction centers are assembled in single photosynthetic machinery. Three- or four photon reactions are feasible if the electron transfer chain is elongated linearly without short circuit. Multi-photon reactions could also be established by modifying electron transfer pathway without increasing the reaction centers. As seen in the high- and low-potential chains in the Q-cycle[47,48] of the cytochrome $b_6f$ complex (an electron transporter between P680 and P700), a two-electron transfer reaction can drive an energetically uphill electron transfer reaction by coupling it with an energetically downhill reaction. By expending one of two electrons during the electron transfer, a two-photon reaction may functionally behave as a three- or four-photon reaction to use NIR radiation (this hypothesis should be explored in future work). A red-edge position at around 900 or 1,100 nm may be an important biosignature for habitable planets around very old M dwarfs.

We thus conclude that future missions should prepare to record surface spectra of habitable exoplanets at wavelengths from shorter than 700 nm to longer than 1,100 nm so that they have the capability to address the possibility that the red-edge position may change as the host M-dwarf ages.

**Methods**

- Estimating the PFD spectrum on the land surface and underwater

The standard terrestrial solar spectral irradiance table, ASTM G173, was obtained online (http://rredc.nrel.gov/solar/spectra/am1.5/). The standard spectral irradiance (W m$^{-2}$ nm$^{-1}$) at the top of the atmosphere (AM0) and that passing through a 1.5 air-mass atmosphere (AM1.5 direct spectrum) were converted to PFD spectra (μmol photon m$^{-2}$ s$^{-1}$ nm$^{-1}$). The solar PFD spectrum from 2.7 billion years ago was obtained online (http://depts.washington.edu/naivpl/content/models/solarflux). Atmospheric transmittance was assumed to be the same as the present (AM1.5), except that the ozone ($O_3$) contribution (total column equivalent of 3.4 mm, 233 K) was subtracted. Underwater PFD spectra at various depths were calculated from the absorption spectrum of distilled water provided by Hale and Querry.[35] The spectral irradiance of AD Leo, provided by Segura et al.,[5] was normalized to have an integrated (120–10,000 nm) power of 1,366 W m$^{-2}$, which is equivalent to the power of solar irradiance at the top of the atmosphere on the Earth. Atmospheric and water transmittances on the hypothetical habitable planet around AD Leo were assumed to be identical to those on the present Earth.

- Calculating the integrated PFD and the energy yield of the reaction center

The integrated PFD for a given wavelength range was approximated by summing the spectral PFD (µmol photon $m^{-2}$ $s^{-1}$ $nm^{-1}$) over the range ($\sum$PFD, µmol photon $m^{-2}$ $s^{-1}$). The energy gain from a single photochemical reaction was calculated from the excitation wavelength of the reaction center ($\lambda$RC) as 1,240/$\lambda$RC (eV) or 1.986 × $10^{-16}$/$\lambda$RC (W s). Assuming 100% quantum efficiency, the total energy input for the reaction center was calculated as $\sum$PFD × 119.6/$\lambda$RC (W $m^{-2}$).

Acknowledgements

This research was supported by the NINS Program for Cross-Disciplinary Study. K.T. and J.M acknowledge support by JSPS KAKENHI Grant Number 16H06553. N.K. and N.N. acknowledge support from a Grant-in-Aid for Challenging Exploratory Research (JSPS KAKENHI Grant Numbers 16K13791 and 16K17671) and a Grant-in-Aid for Scientific Research (A) (JSPS KAKENHI Grant Number 25247026), respectively. M.T. acknowledges support by JSPS KAKENHI Grant Number 15H02063.

Author contributions

K.T. and N.N. conceived this study. K.T. performed the calculations. K.T., N.N., J.M., M.T., and N.K. wrote and reviewed the manuscript.

Competing financial interests

The authors declare no competing financial interests.

Table Captions

Table 1 | Lighting conditions on the Earth and a habitable planet around AD Leo. Parameters for EFD (a) and PFD (b) were obtained from spectra shown in Supplementary Figure 1. Integrated PFDs were calculated for a fixed bandwidth at 300 nm (c) or a variable bandwidth (d). Integrated PFDs increased as bandwidth increased and were not saturated. Total energies obtained at the reaction centers (RCs) were calculated from the integrated PFDs and the RC wavelength, as shown in sections e and f.

Table 1

| | Sun/Earth | | | | AD Leo | | | |
|---|---|---|---|---|---|---|---|---|
| | Surface | Underwater | | | Surface | Underwater | | |
| | | 0.1 m | 1 m | 10 m | | 0.1 m | 1 m | 10 m |

a) Energy Flux Density

| | | | | | | | | |
|---|---|---|---|---|---|---|---|---|
| Peak wavelength (nm) | 531 | 531 | 495 | 495 | 1005 | 809 | 656 | 539 |
| Peak value (W m$^{-2}$nm$^{-1}$) | 1.435 | 1.423 | 1.344 | 0.802 | 1.138 | 0.617 | 0.227 | 0.082 |
| Relative value at 700 nm | 0.81 | 0.71 | 0.22 | 0.00 | 0.39 | 0.63 | 0.49 | 0.00 |

b) Photon Flux Density

| | | | | | | | | |
|---|---|---|---|---|---|---|---|---|
| Peak wavelength (nm) | 669 | 669 | 531 | 495 | 1005 | 809 | 656 | 539 |
| Peak value ($\mu$mol m$^{-2}$s$^{-1}$nm$^{-1}$) | 7.290 | 6.660 | 5.875 | 3.318 | 9.564 | 4.174 | 1.247 | 0.368 |
| Relative value at 700 nm | 0.93 | 0.89 | 0.29 | 0.00 | 0.27 | 0.54 | 0.52 | 0.00 |

c) Best band range (nm)

| | | | | | | | | |
|---|---|---|---|---|---|---|---|---|
| Best band range (nm) | 578–878 | 441–741 | 397–697 | 349–649 | 985–1285 | 606–906 | 416–716 | 365–665 |
| Max value ($\mu$mol m$^{-2}$s$^{-1}$) | 1965 | 1730 | 1240 | 408 | 2205 | 659 | 164 | 35 |
| Relative value at 400–700 nm | 0.88 | 0.97 | 1.00 | 0.98 | 0.12 | 0.38 | 0.98 | 0.99 |
| 700–900 nm | 0.99 | 0.75 | 0.26 | 0.00 | 0.55 | 1.00 | 0.49 | 0.00 |
| 900–1100 nm | 0.80 | 0.17 | 0.00 | 0.00 | 0.96 | 0.47 | 0.01 | 0.00 |

d) Integrated PFD: variable bandwidth

| | | | | | | | | |
|---|---|---|---|---|---|---|---|---|
| Max band range (nm) | 400–1400 | 400–1400 | 400–1400 | 400–1400 | 400–1400 | 400–1400 | 400–1400 | 400–1400 |
| Max value ($\mu$mol m$^{-2}$s$^{-1}$) | 4806 | 2373 | 1259 | 401 | 4364 | 806 | 169 | 35 |
| Relative value at 400–700 nm | 0.36 | 0.70 | 0.98 | 1.00 | 0.06 | 0.31 | 0.94 | 1.00 |
| 400–900 nm | 0.63 | 0.99 | 1.00 | 1.00 | 0.30 | 0.94 | 1.00 | 1.00 |
| 400–1100 nm | 0.82 | 1.00 | 1.00 | 1.00 | 0.64 | 1.00 | 1.00 | 1.00 |

e) Total Energy at RC: bandwidth 300 nm

| | | | | | | | | |
|---|---|---|---|---|---|---|---|---|
| Best band range (nm) | 447–747 | 411–711 | 369–669 | 277–577 | 797–1097 | 594–894 | 407–707 | 282–582 |
| Max value (W m$^{-2}$) | 302 | 286 | 216 | 83 | 229 | 88 | 28 | 7 |
| Relative value at 400–700 nm | 0.98 | 1.00 | 0.98 | 0.82 | 0.20 | 0.49 | 0.99 | 0.84 |
| 700–900 nm | 0.86 | 0.60 | 0.20 | 0.00 | 0.70 | 1.00 | 0.38 | 0.00 |
| 900–1100 nm | 0.57 | 0.11 | 0.00 | 0.00 | 1.00 | 0.38 | 0.00 | 0.00 |

f) Total Energy at RC: variable bandwidth

| | | | | | | | | |
|---|---|---|---|---|---|---|---|---|
| Best band range (nm) | 400–1110 | 400–846 | 400–698 | 400–577 | 400–1342 | 400–912 | 400–709 | 400–582 |
| Max value (W m$^{-2}$) | 431 | 317 | 212 | 82 | 388 | 101 | 28 | 7 |
| Relative value at 400–700 nm | 0.69 | 0.90 | 1.00 | 0.84 | 0.12 | 0.42 | 0.99 | 0.84 |
| 400–900 nm | 0.93 | 0.98 | 0.79 | 0.65 | 0.45 | 1.00 | 0.81 | 0.66 |
| 400–1100 nm | 1.00 | 0.81 | 0.65 | 0.53 | 0.78 | 0.86 | 0.66 | 0.54 |

Figure Captions

Figure 1 | Lighting conditions on a hypothetical habitable planet around an M-dwarf and the evolution of photosynthesis. Ovals and arrows outline the flow of evolutionary paths from a two-photon reaction using visible radiation (Vis-Vis) to a two-color reaction using visible and NIR radiation in separate reaction centers (Vis-NIR). The area graph on the left side shows the visible-radiation/NIR-radiation ratio on the land surface and underwater at different depths.

Figure 2 | Light intensity changes in shallow water. Solar (panel a) and AD Leo (panel b) radiation intensities obtained for shallow water are plotted against water depth. Black squares, red circles, and blue triangles represent the integrated PFDs in the ranges of 400–700, 400–900, and 400–1,100 nm, respectively.

Figure 3 | State transitions in hypothetical photosynthesis using NIR radiation in shallow water. Ovals show the reaction centers excited by visible ($P_{vis}$) and NIR ($P_{NIR}$) radiation. Light gray circles show light-harvesting antenna absorbing visible radiation. Dark gray circles show light-harvesting antenna absorbing visible and NIR radiation. Shuttling model: When the cell descends in the water, a part of the visible-radiation-harvesting antenna moves from $P_{vis}$ to $P_{NIR}$ (shown in the left scheme). Quenching model: When the cell rises towards the surface, part of the visible-radiation-harvesting antenna detaches from $P_{vis}$ and transitions to a quenching state (shown as open circles at the upper right).

Figure 4 | Light-harvesting antenna size regulation in shallow water. The changes in the light-harvesting antenna sizes for Earth-type two-photon reactions (P680

and P700) and heterogeneous reaction centers (P700 and P900) were calculated for the state transition model. Panels a and d show the antenna size changes for P680 (black squares) and P700 (red circles) under solar radiation. Panels b and e show the antenna size changes for P680 (black squares) and P700 (red circles) under AD Leo's radiation. Panels c and f show antenna size regulation for P700 (black squares) and P900 (red circles) under AD Leo's radiation. Effective antenna size (Abs) of the reaction centers is plotted against water depth. In panels a, b, and c, antenna size regulation by the shuttling model was calculated as the water depth increased from 0 to 10 m. In panels d, e, and f, antenna size regulation by the quenching model was calculated as water depth decreased from 1 to 0 m (antenna size regulation by the shuttling model was also calculated as water depth increased from 1 to 10 m).

Figure 5 | Regulation of the electron transfer reaction in shallow water. The changes in LEF (black squares) and CEF (red circles) as a result of state transitions are plotted as a function of water depth. Panels a−f correspond to panels a−f in Figure 4. CEF and LEF were calculated from the antenna size and light intensity.

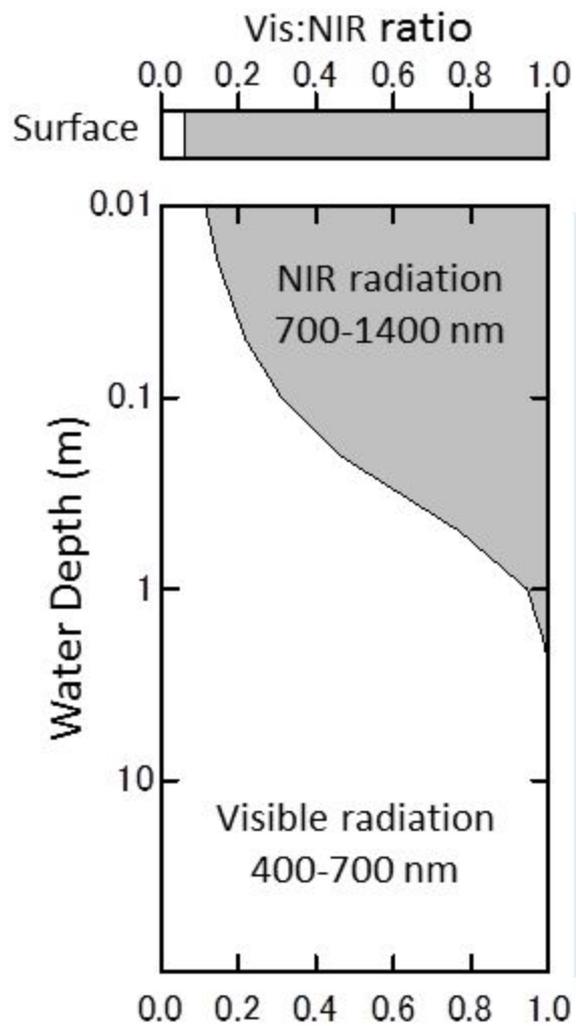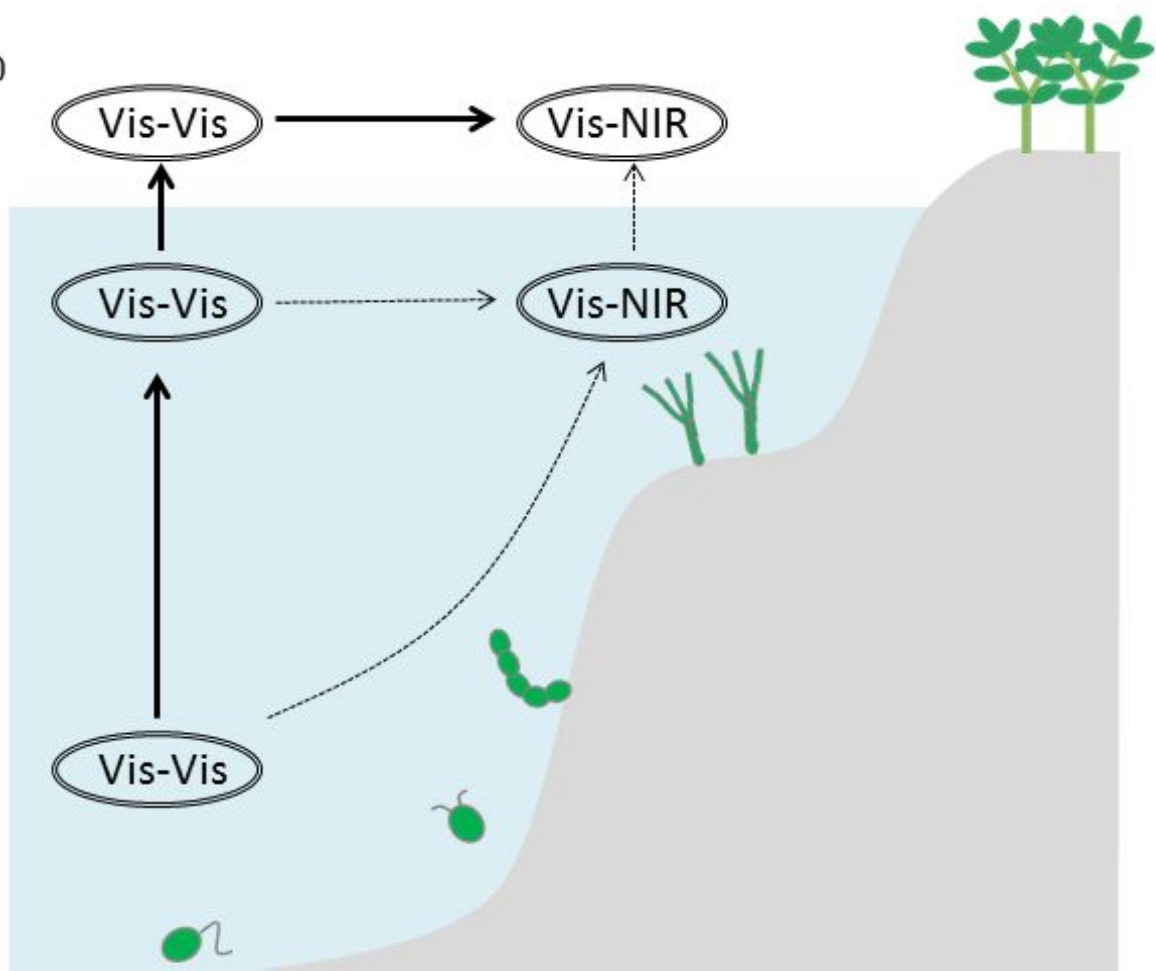

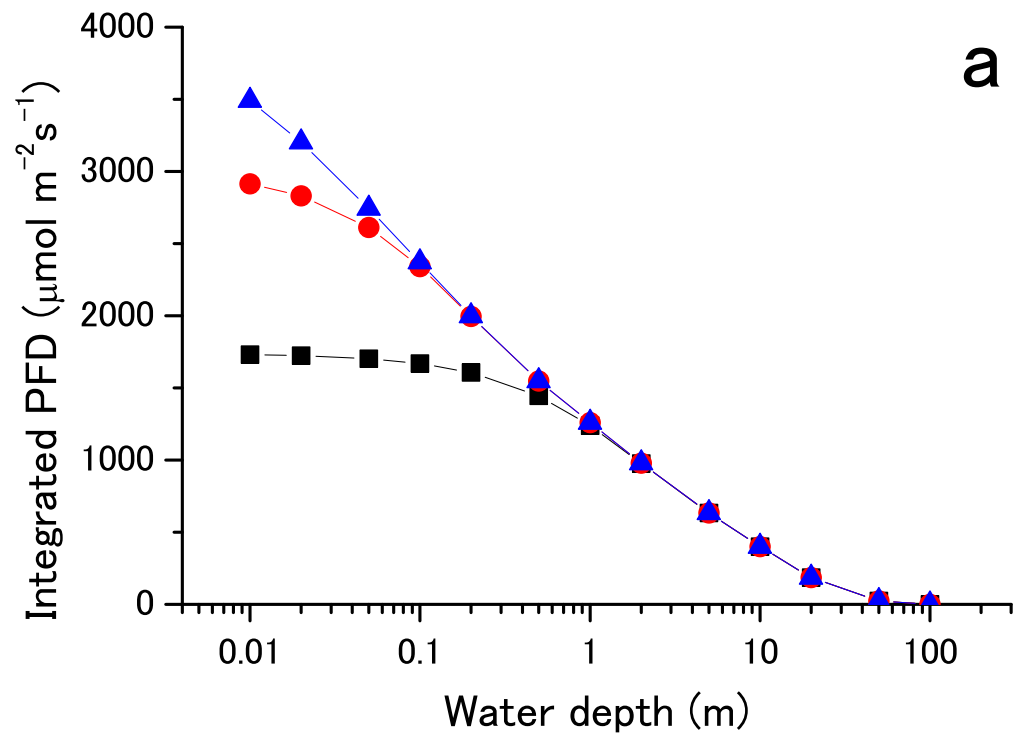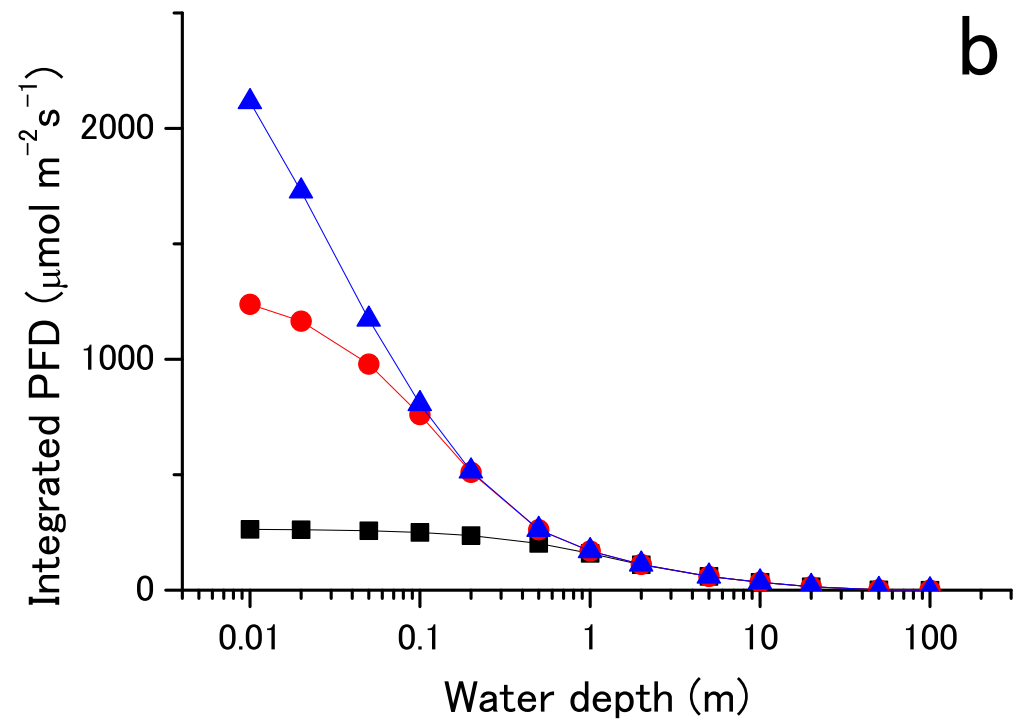

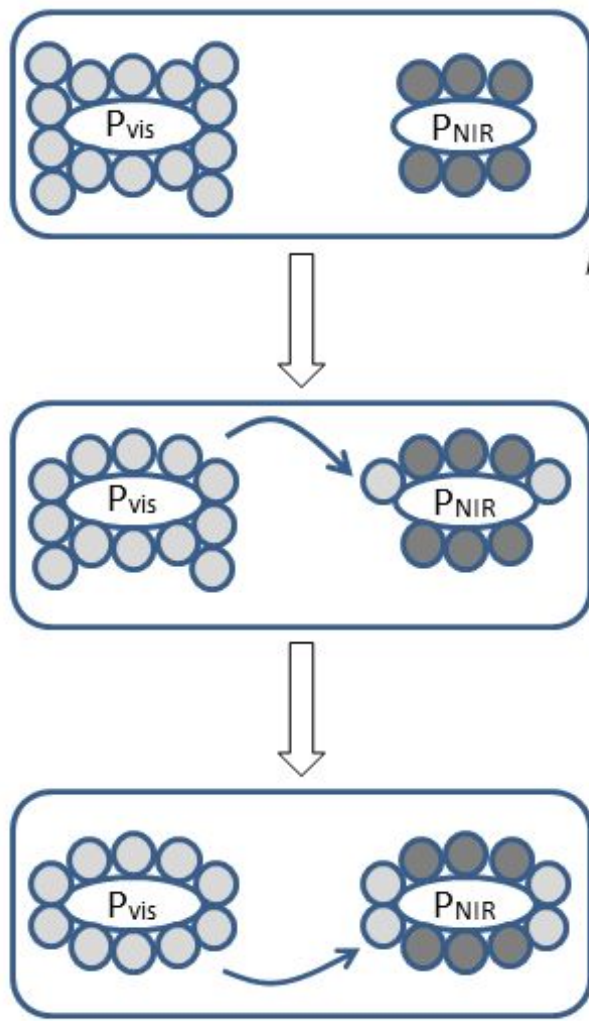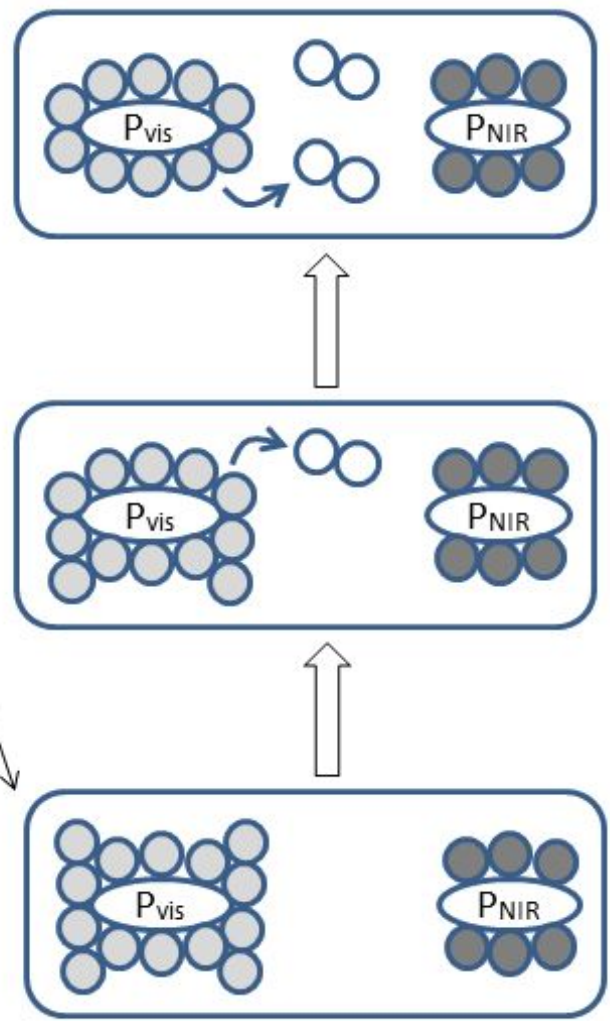

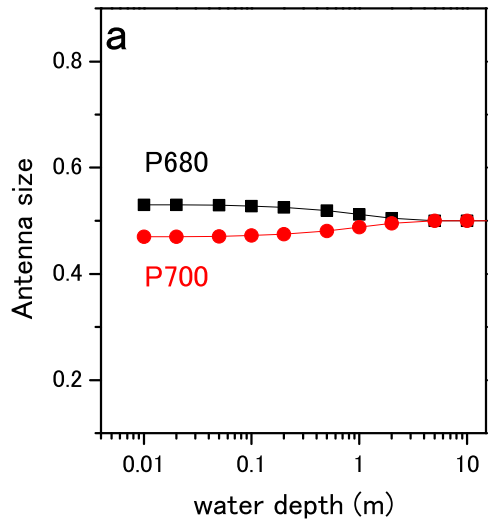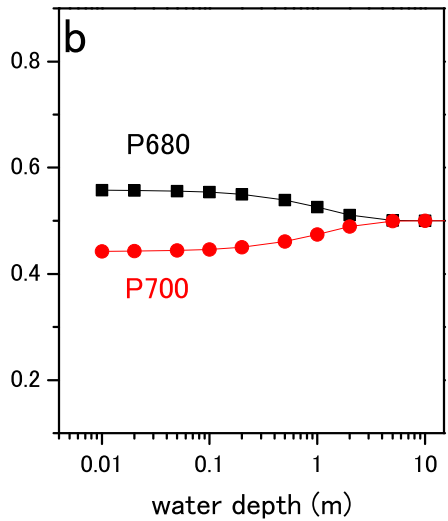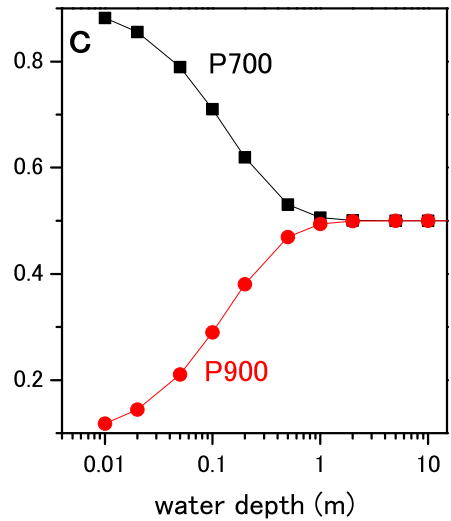
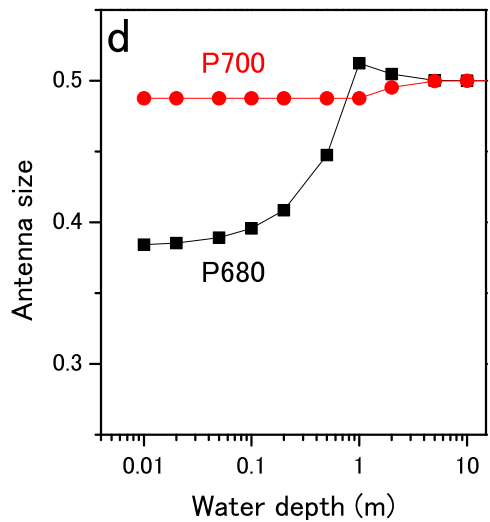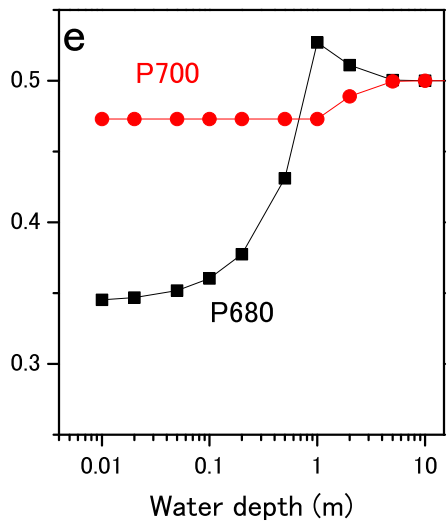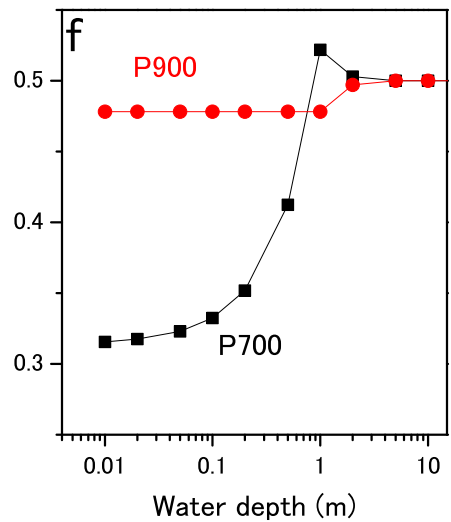

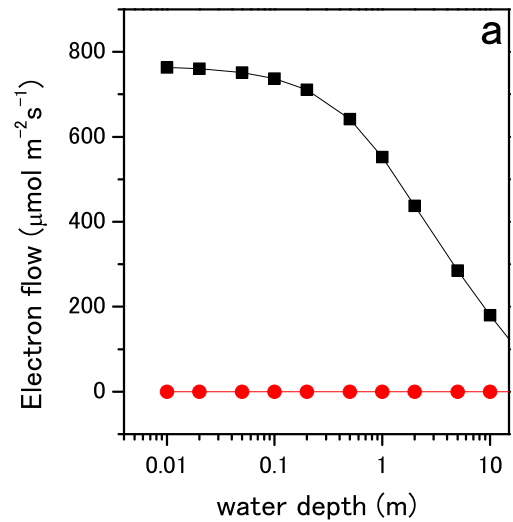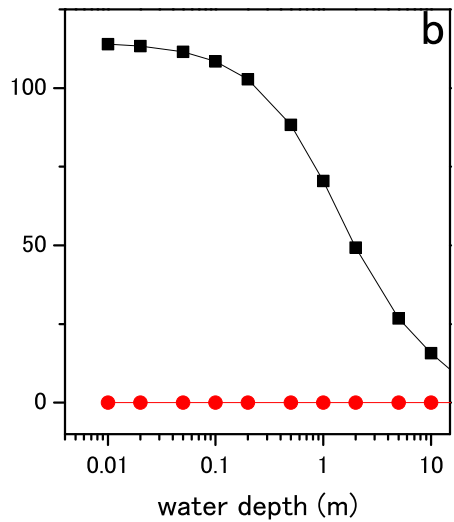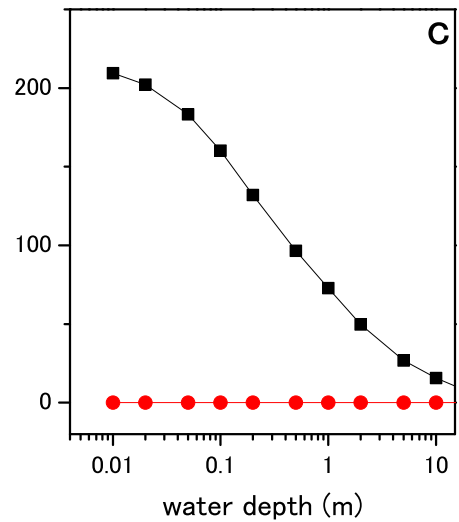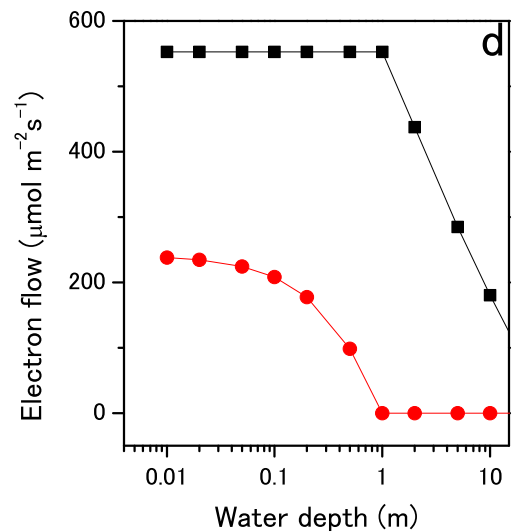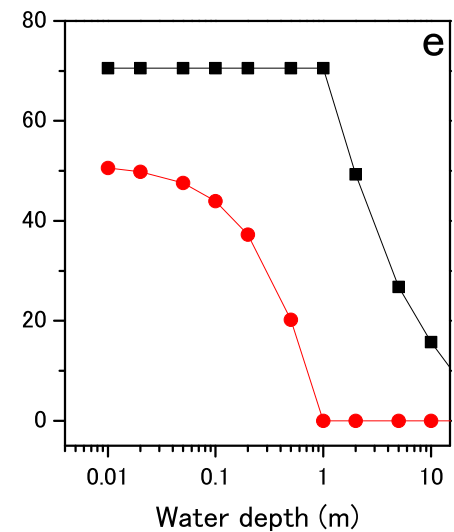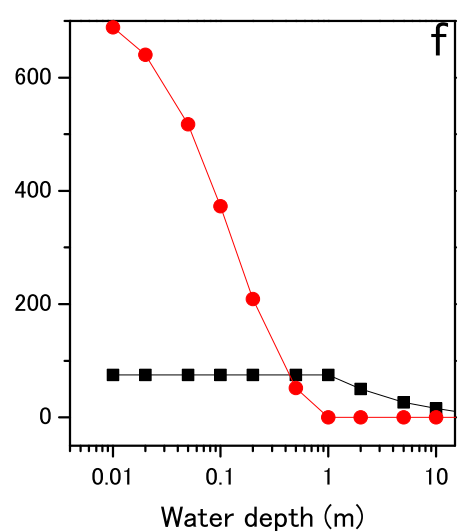